# Two-Sided Impact of Water on the Relaxation of Inner-Valence Vacancies of Biologically Relevant Molecules

Anna D. Skitnevskaya,* Kirill Gokhberg, Alexander B. Trofimov, Emma K. Grigoricheva, Alexander I. Kuleff, and Lorenz S. Cederbaum

**ABSTRACT:** After ionization of an inner-valence electron of molecules, the resulting cation-radicals store substantial internal energy which, if sufficient, can trigger ejection of an additional electron in an Auger decay usually followed by molecule fragmentation. In the environment, intermolecular Coulombic decay (ICD) and electron-transfer mediated decay (ETMD) are also operative, resulting in one or two electrons being ejected from a neighbor, thus preventing the fragmentation of the initially ionized molecule. These relaxation processes are investigated theoretically for prototypical heterocycle−water complexes of imidazole, pyrrole, and pyridine. It is found that the hydrogen-bonding site of the water molecule critically influences the nature and energetics of the electronic states involved, opening or closing certain relaxation processes of the inner-valence ionized system. Our results indicate that the relaxation mechanisms of biologically relevant systems with inner-valence vacancies on their carbon atoms can strongly depend on the presence of the electron-density donating or accepting neighbor, either water or another biomolecule.

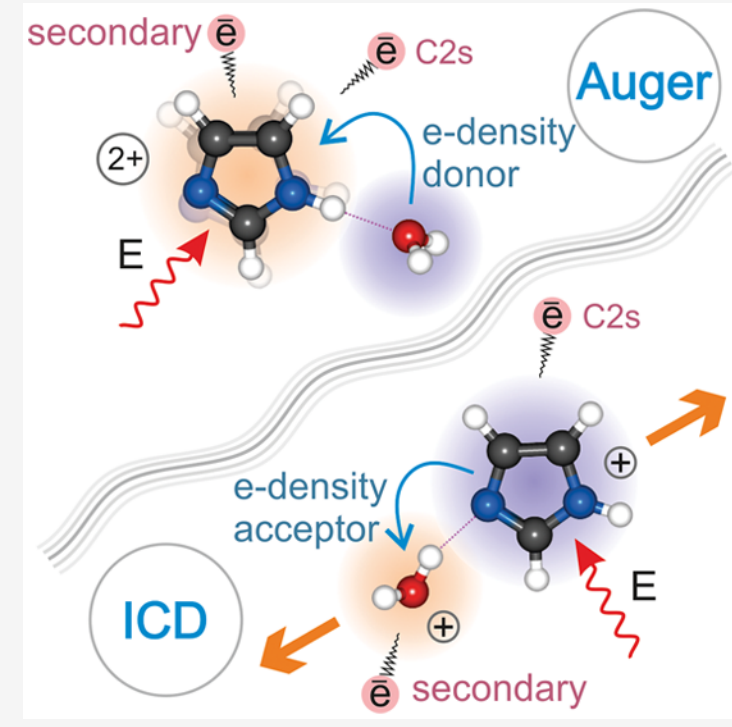
secondary e
e C2s
Auger
2+
e-density
donor
E
e C2s
e-density
acceptor
ICD
E
e secondary

Extreme (from 10 up to 124 eV) ultraviolet (XUV) irradiation of biological matter can cause inner-valence ionization, producing excited cation-radicals and depositing large amount of energy which can be released via a number of competing relaxation channels. In the case in which the populated excited cationic state is higher in energy than the lowest double-ionization potential, the system can relax by emitting a secondary electron. A prominent example of such a pathway is the Auger decay in which the initial inner-valence vacancy is filled by an outer-valence electron and the excess energy is used to emit another outer-valence electron form the same molecule[1−3] (Figure 1). The resulting two vacancies are thus located on the initially ionized molecule and can often cause a bond breaking or system fragmentation.[4,5]

Apart from some modifications in the energetics, the Auger process and its efficiency are very little affected by the chemical environment of the ionized molecule.[6,7] The presence of weakly bound neighbors, as is the case of biomatter and solutions, opens a whole plethora of very efficient, nonlocal electronic decay processes that involve electrons from the systems in the environment.[8,9] These processes also terminate by creating two vacancies, which are, however, distributed on different molecules. This might substantially reduce the probability for fragmentation of the initially ionized molecule and thus might appear as a protective mechanism in biomatter. The created radical cations in the environment, however, are usually highly reactive and can induce further structural changes and damage to the system purely chemically. The competition between these local and nonlocal decay mechanisms, as well as the follow-up complex dynamics, which are still far from being fully understood, determines the degree of damage induced by irradiating the system with XUV light. The detailed knowledge of these mechanisms is, therefore, the key to understanding the radiation damage in biosystems and to eventually develop protective strategies.

The presence of the environment of the initially ionized molecule can open several decay channels even in cases when the local Auger decay is energetically forbidden. Typically, the most efficient of those is the interatomic or intermolecular Coulombic decay (ICD),[8] in which the energy released from the relaxation of the initially ionized fragment is transferred to a neighboring atom or molecule that uses it to emit one of its electrons. As a result, two cation-radical monomers repelling each other and a low-energy electron are produced (Figure 1). The ICD was first predicted theoretically by Cederbaum and co-workers in 1997[10] and later confirmed experimentally in a large variety of atomic and molecular clusters.[11−19] The process is proved to take place on an ultrashort time scale (usually in the femtosecond regime),[20,21] and contrary to the

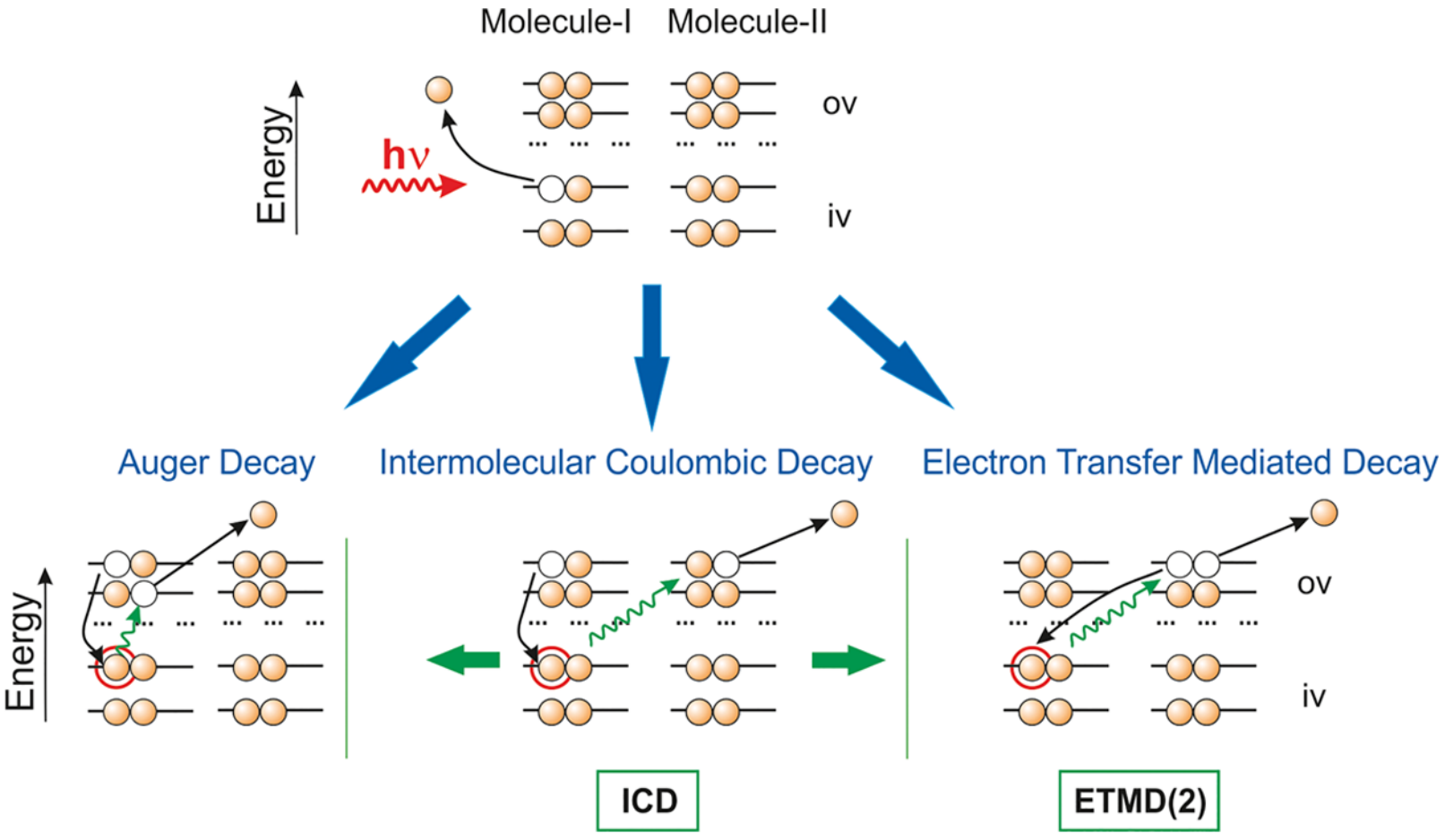

**Figure 1.** Schematic representation of the different electronic decay processes initiated by an inner-valence (iv) ionization of one of the molecules: local relaxation, or Auger decay (left), and nonlocal relaxation, ICD (middle) and ETMD processes (right).

local decay mechanisms, its efficiency increases with the number of neighbors in the environment.[13,22]

A usually slower, but in many cases the only possible decay mechanism, is the electron-transfer mediated decay (ETMD).[23−25] In this process, an electron from a neighboring atom or molecule fills the initial vacancy causing the emission of another electron from the environment. As a result, two vacancies are formed on the neighboring fragment(s) while the initially ionized species becomes neutral (Figure 1).

Being nonlocal channels, ICD and ETMD are considered to cause ionization of biomolecules when the environment is initially ionized, and *vice versa* to distribute the internal energy to the environment when the biomolecule is ionized by the initial irradiation. It was shown both theoretically and experimentally that ICD is possible after inner-valence ionization of heteroatoms in simple molecules[26,27] and also after ionization of C 2s in larger systems, especially in $\pi$-conjugated ones.[28,29] It was also shown that the entire inner-valence ionization energy region is open for ICD in comparatively large biologically relevant aromatic molecules.[28] As the biological molecules are mostly solvated in water, the water molecule is considered as the universal partner in the ICD or ETMD after inner-valence ionization.[8] It was shown that inner-valence ionization of the water molecule will be followed by ICD involving a neighboring organic molecule.[30,31] The same was also recently demonstrated to be the case for liquid water[19] and for the tetrahydrofuran·$H_2O$ system (a model of solvated deoxyribose fragment in DNA).[32] For the latter system, it was also theoretically predicted that the inner-valence ionization of water can initiate an ETMD with the tetrahydrofuran, but this process could not be resolved experimentally as it occurs in the same energy region and produces the same final states as the Auger decay of oxygen $2s^{-1}$ states of tetrahydrofuran.

Considering the case of biosystems, it is hard to overestimate the role of hydrogen bonding. It is known that in the event of ionization, intermolecular hydrogen bonding, being donor−acceptor by its nature, ensures stabilization or destabilization of the vacancies which are formed. As shown for a number of systems,[26,30] the Auger final states with two holes located on one molecule are significantly stabilized (accessible with lower energy) by the electron-density donating neighbor and destabilized by an electron-density acceptor. For systems of small molecules, which are usually the object of both theoretical and experimental studies of electronic decay processes, the lowest Auger or ETMD final states are often energetically higher than the lowest ICD states. However, for a number of molecule−water complexes it was shown[26] that the corresponding energy gap is larger if the water molecule acts as electron-density acceptor and smaller if water acts as the electron-density donor stabilizing the positive charges on the neighboring molecule. The aim of the present work is to study the role of the donor−acceptor interactions with water in the case of inner-valence ionization for systems that mimic electron-density-rich biomolecules. As such, we took nitrogen-containing heterocycles: imidazole, pyrrole, and pyridine (see Scheme 1).

**Scheme 1. Nitrogen-Containing Heterocycles**

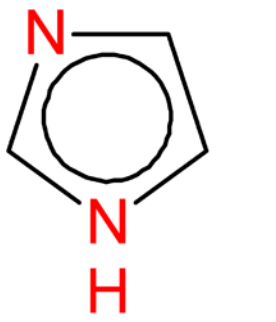

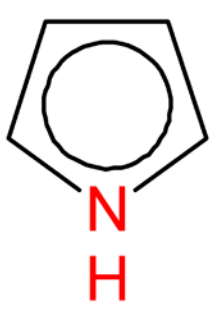

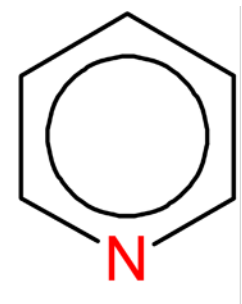

This choice of relatively small but representative model systems is motivated by the possibility to perform our analysis with the help of high-precision *ab initio* approaches, but also to offer suitable systems for future experimental studies, where the basic mechanisms can be well disentangled.

Structures of heterocycle−water clusters were considered in numerous works.[33−38] We thus used these results as a guide to choose the most thermodynamically stable complexes for the calculations of the ionization (IP) and double ionization (DIP) spectra needed for the further analyses. All geometries, normal modes, and basis-set superposition error (BSSE) corrections for the structures considered in this work were obtained using MP2/aug-cc-pVDZ level of theory, as implemented in the Gaussian16 program suite.[39]

The preferable ways a water molecule binds to the heterocycles are shown in Figure 2. One can see that there are two principal types of water coordination: (i) by hydrogen to the lone pair of nitrogen and (ii) by oxygen to one of the hydrogens of the heterocycle. Therefore, with respect to the heterocycle, water acts as an electron-density acceptor ($a$-$H_2O$) or electron-density donor ($d$-$H_2O$) (see Figure 2).

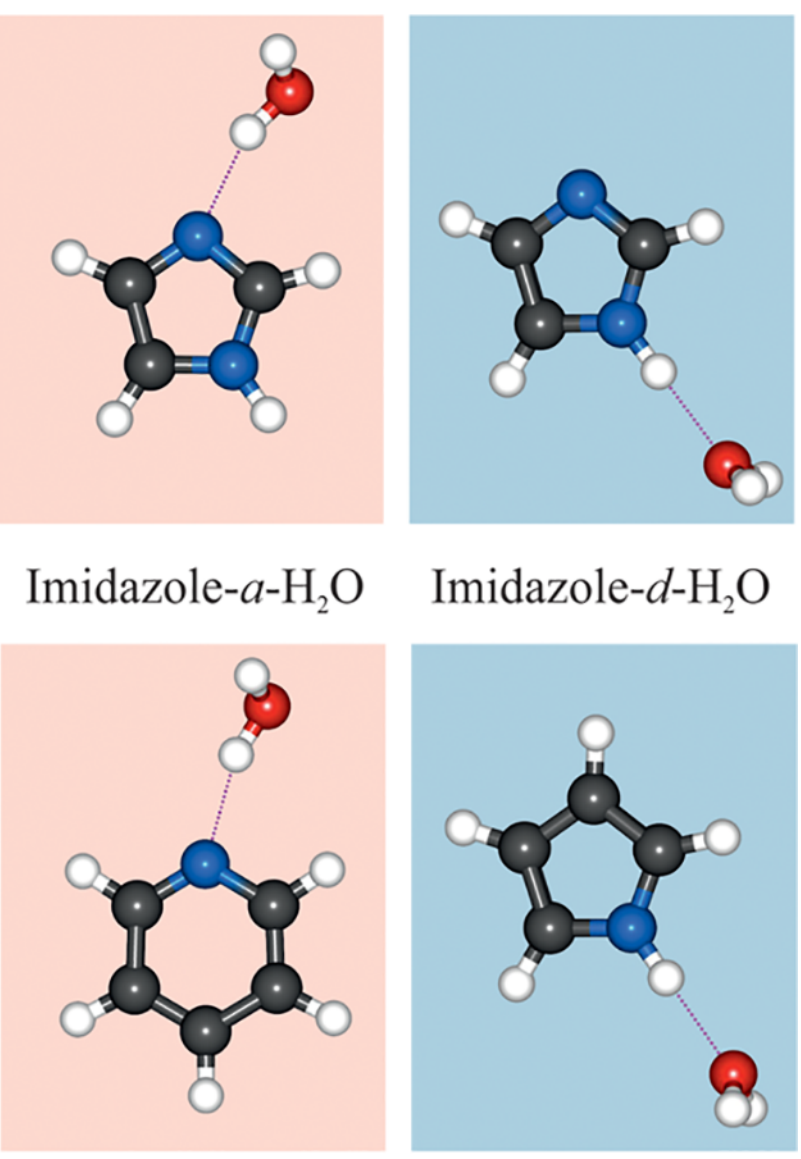

**Figure 2.** Structures of the most stable complexes of imidazole, pyridine, and pyrrole with water. The indexes *a* and *d* indicate whether water acts as an electron-density acceptor or electron-density donor, respectively.

An important feature of the imidazole molecule is the two nitrogen atoms differently incorporated in the aromatic system, one of which is of pyrrole type, the other one of pyridine type. According to refs 40−43, O−H···N bonding in the imidazole-$a$-$H_2O$ structure is more likely to form than O···H−N in imidazole-$d$-$H_2O$ (Figure 2). Our calculations show that the corresponding energy difference is 0.6 kcal/mol, which is consistent with previous results.[40−43] Nevertheless, both variants are experimentally observed[43] and their relative abundance in some cases is found to be up to 1:1.[41] We, therefore, consider here both structures. This also allows us to directly compare the effect of the two types of hydrogen bonding on the local and nonlocal decay processes.

In most of the works on pyridine−water system,[37,38] only O−H···N bonding type is considered. According to our results, pyridine-$a$-$H_2O$ (Figure 2) is substantially lower in energy (up to 3.0 kcal/mol) compared to all alternative structures. We will, therefore, further consider only this one.

The bonding nature of the nitrogen atom in pyrrole is different from pyridine, as the nitrogen provides two electrons to the pentahedral aromatic $\pi$-system and one electron to form a $\sigma$-bond with the hydrogen atom. Thus, a hydrogen bond of O···H−N type is formed, providing pyrrole-$d$-$H_2O$ structure (Figure 2)[35,36] with water acting as an electron-density donor with respect to the pyrrole ring. According to our calculations, the other possible structure with the water hydrogen bonded to the pyrrole $\pi$-system (pyrrole-$a$-$H_2O$) is higher in energy by 1.0 kcal/mol. Moreover, to the best of our knowledge, only pyrrole-$d$-$H_2O$ structure is obtained under experimental conditions, and therefore, only this structure was considered in this work.

As was mentioned above, the system can relax from the singly to doubly ionized state losing an additional electron if the energy of the dicationic state is lower than that of the initial cation. To determine the open channels, we can thus compare the single- and double-ionization spectra of the heterocycle−water complexes. To do so, we computed the IP and DIP spectra of imidazole-$a$-$H_2O$ and imidazole-$d$-$H_2O$ molecular pairs, as well as of pyridine-$a$-$H_2O$ and pyrrole-$d$-$H_2O$ (Figures 3 and 4). Those spectra were calculated using the third-order (ADC(3)) and second-order (ADC(2)) algebraic diagrammatic construction scheme for obtaining the one- and two-particle propagators, respectively, in combination with cc-pVDZ bases sets (see Computational Methods for further details). The obtained spectra are depicted in Figures 3 and 4.

It can be seen that a large number of states in the inner-valence region lie in the cationic continuum, i.e., are above the lowest dicationic state, and thus can relax by emitting an electron through one of the possible decay mechanisms and form a dication. To distinguish the possible channels, the different types of dicationic states are marked by different colors in Figures 3 and 4 (middle panels). The states with two holes localized on different molecules are plotted in cyan and correspond to ICD final states, regardless of which molecule in the pair was initially ionized. Dications having both holes on the heterocycle (states plotted in orange) can be formed either through an Auger decay channel, if the initial vacancy is located on the heterocyclic molecule, or through an ETMD channel, if the initial vacancy is located on the water. As we will discuss later, the states with two vacancies located on the water molecule are not accessible in the energy range considered in this work.

The energy difference between singly and doubly ionized states corresponds to the energy of the emitted secondary electrons, which makes possible the modeling of the secondary electron spectra, as described in ref 30 and in Computational Methods. If we assume that all cationic states up to about 35 eV are initially populated, which covers all inner-valence ionizations, we obtain the electron spectra that are depicted in Figures 3 and 4, upper panels. The contributions of the different channels are also shown, indicating that ICD and ETMD electrons can be expected to contribute substantially to the secondary electron spectra.

Despite the overall similarity of the IP spectra of imidazole-$a$-$H_2O$ and imidazole-$d$-$H_2O$, there are substantial differences that can be attributed to the acceptor or donor behavior of water toward imidazole. When water is bonded in an electron-density donating position, it stabilizes the vacancies on imidazole, resulting in lowering of the energies of the corresponding ionic states and *vice versa* for the accepting position. Consequently, the first ionization potential of the two complexes corresponding in both cases to the removal of an electron from the HOMO of imidazole appears with an energy difference of 0.8 eV (Figure 3). Moreover, the states corresponding to the water ionization, depicted with green lines in the lower panels of Figure 3, are systematically and substantially lower in energy for the imidazole-$a$-$H_2O$ compared to the imidazole-$d$-$H_2O$, due to the ability (or the disability, respectively) of imidazole to screen the vacancies formed on the water. The largest difference is 2.5 eV and appears for the lowest IP of water ($1b_1^{-1}$), while the smallest energy shift is observed for the third IP of water ($1b_2^{-1}$). However, due to the stronger correlation effects present in imidazole-$d$-$H_2O$, the removal of a $1b_2$ electron from the donor water leads to a much richer structure, with large number of satellite states spread over several electronvolts.

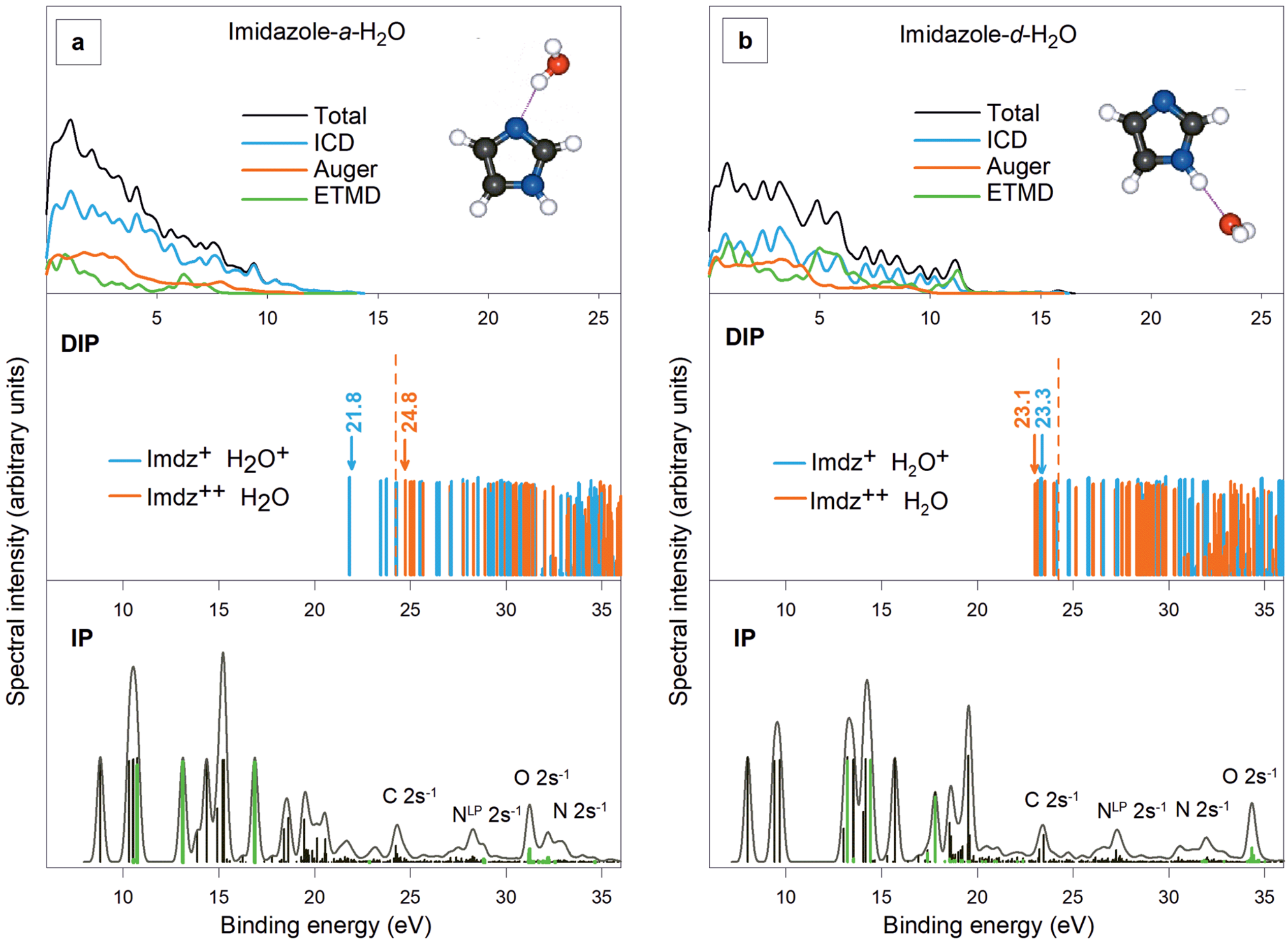

**Figure 3.** Lower panels: Single-ionization spectra (IP) of imidazole−water complexes calculated at ADC(3)/cc-pVDZ level (vertical bars). To account for the vibrational broadening, the spectrum was convoluted with a Gaussian with fwhm of 0.4 eV (solid line envelope). The green bars denote states populated by ionization of water. The bands corresponding to inner-valence vacancies ($2s^{-1}$ states) localized on the different atoms are also marked. Middle panels: Double-ionization spectra (DIP) of imidazole−water complexes calculated at ADC(2)/cc-pVDZ level of theory. States with both vacancies localized on the heterocycle are depicted in orange, and those with one hole on the imidazole and one on water are depicted in blue. The vertical dashed line indicates the position of the lowest double-ionization potential of the isolated imidazole. Upper panels: simulation of the expected total and partial electron spectra, resulting from the decay processes initiated by the population of all single-ionization states up to 35 eV. For further details, see the text.

In the case of the dicationic heterocycle in imidazole-*d*-$H_2O$ system, the water stabilizes the imidazole$^{2+}$ states with rather pronounced energy downshift of about 1 eV relative to the isolated imidazole molecule (the position of the lowest double-ionization potential of the latter is shown as a dashed line in the middle panels of Figure 3). At the same time, the energy of the lowest ICD state is relatively high (compared to imidazole-*a*-$H_2O$), which together results in vanishing of the gap between the lowest ICD and Auger states (Figure 3b). To the best of our knowledge, the possibility to close the energy gap between the lowest ICD and Auger states, due to the donor−acceptor intermolecular interactions, was not reported previously.

To support this result we performed the same calculations for the pyrrole-*d*-$H_2O$ system (Figure 4b), where water preferably coordinates to the nitrogen atom in a donating position, similar to imidazole-*d*-$H_2O$. Here the stabilization of doubly ionized states relative to isolated pyrrole molecule is 1.1 eV, resulting in the first Auger state being 1 eV lower than the first ICD state. This surprising result can be explained by the redistribution of the electron density within the complex. The reduced electronic density around the water makes the removal of a water electron more costly, while the increased electron density around the heterocycle lowers the energy needed for removing additional electrons from the heterocycle. This twofold effect is so strong that it cannot be compensated by the reduced Coulomb repulsion in the case of distributed final holes, making the Auger final state the lowest dicationic state for these complexes. We note that our calculations on furan-*d*-$H_2O$ and thiophene-*d*-$H_2O$ (not shown) confirm this tendency.

In contrast, in the imidazole-*a*-$H_2O$ molecular pair, the electron-density accepting behavior of water causes destabilization of the states of doubly charged imidazole and thus an upshift of the energies of the final Auger states. The latter, together with the lowering of the final ICD states, results in a 3 eV gap between the lowest ICD and Auger states (Figure 3a). Similar behavior is obtained for the pyridine-*a*-$H_2O$ system, where the water is preferably coordinated to the nitrogen atom and acts as an electron-density acceptor (Figure 4a). Consequently, the Auger final states are slightly destabilized compared to the isolated pyridine and the first one appears 1.5 eV higher than the lowest state of ICD type. To examine the trend, we also computed the DIP spectrum of the pyrrole-*a*-$H_2O$ structure with one of the water hydrogens oriented toward the $\pi$-density of pyrrole. Such a structure is not very stable and less probable, compared to pyrrole-*d*-$H_2O$, but it can confirm our observation that if the water molecule acts as an electron-density acceptor the lowest dicationic state has its holes distributed over the two molecules. Indeed, as in the two other considered cases, the lowest dicationic state of pyrrole-*a*-$H_2O$ is of the ICD type, while the lowest Auger state is 0.6 eV higher. In the case of electron-density accepting behavior of water, the lowering of the final ICD states results in much more intense ICD-electron spectrum, as can be seen by comparing imidazole-*a*-$H_2O$ and imidazole-*d*-$H_2O$ (Figure 3). Due to the different heterocycles involved, such direct comparison cannot be made for pyridine-*a*-$H_2O$ and pyrrole-*d*-$H_2O$ systems.

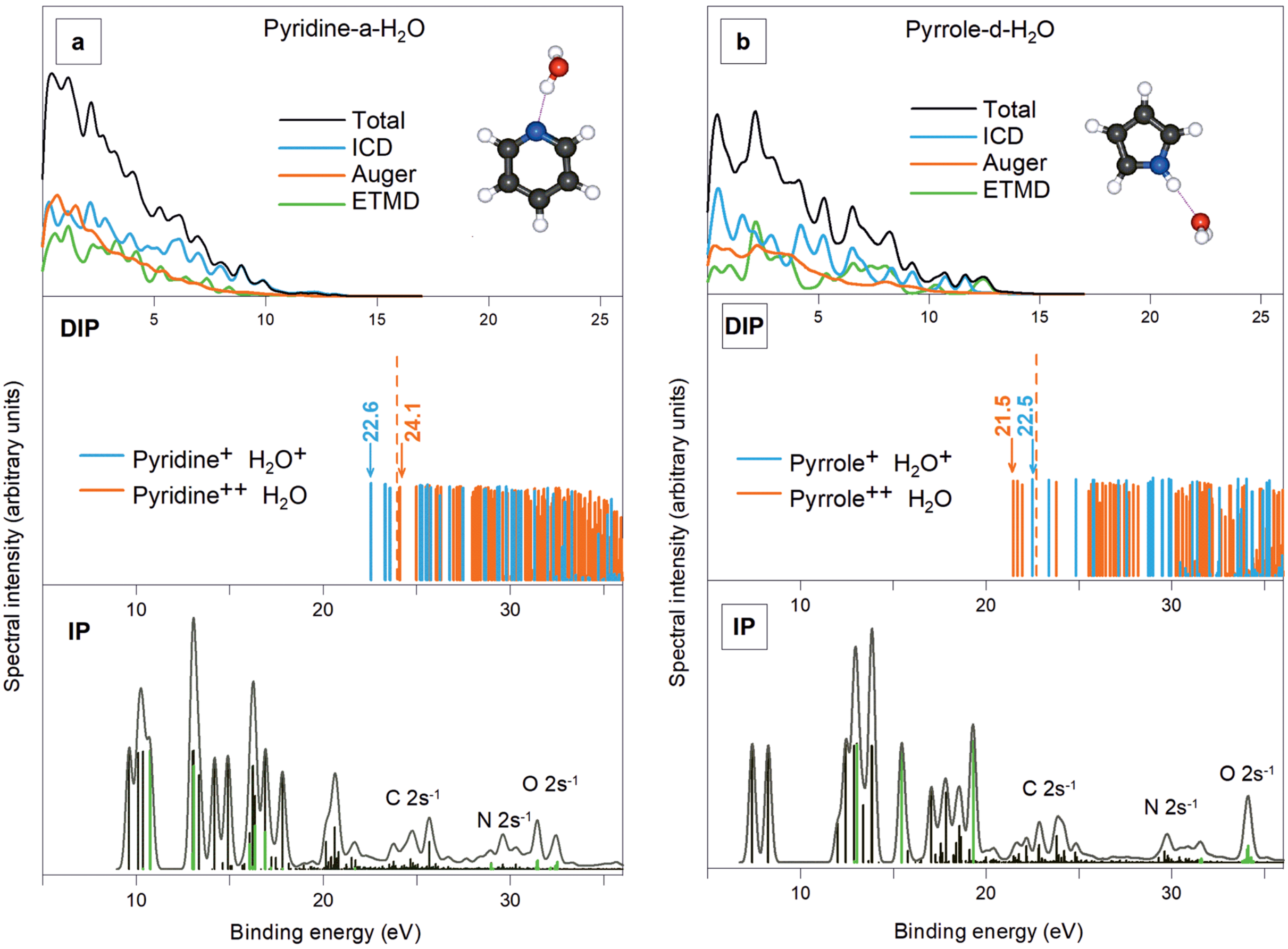

**Figure 4.** Lower panels: Single-ionization spectra of pyridine−water and pyrrole−water complexes calculated at ADC(3)/cc-pVDZ level (vertical bars). To account for the vibrational broadening, the spectrum was convoluted with a Gaussian with fwhm of 0.4 eV (solid line envelope). The green bars denote states populated by ionization of water. The bands corresponding to inner-valence vacancies ($2s^{-1}$ states) localized on the different atoms are also marked. Middle panels: Double-ionization spectra of pyridine−water and pyrrole−water complexes calculated at ADC(2)/cc-pVDZ level of theory. States with both vacancies localized on the heterocycle are depicted in orange, and those with one hole on the heterocycle and one on water are depicted in blue. The vertical dashed line indicates the position of the lowest double-ionization potential of the isolated pyridine and pyrrole, respectively. Upper panels: simulation of the expected total and partial electron spectra, resulting from the decay processes initiated by the population of all single-ionization states up to 35 eV. For further details, see the text.

In isolated imidazole, pyrrole, and pyridine cation-radicals, the energetically deepest inner-valence ionized states correspond to ionization out of the N 2s orbitals. In all cases, the system can then undergo an Auger decay, which will most likely cause molecular fragmentation. In the presence of water, regardless of its position, additional ICD channels open for such initial states, resulting in vacancy separation and thus providing routes for the heterocycle ring to avoid fragmentation.

Slightly lower in the energy, in the region from about 22 to 26 eV, the deepest C $2s^{-1}$ states appear which cannot decay electronically in the case of isolated heterocycles. The electron-density donating position of water in the vicinity of pyrrole results in lowering of the pyrrole$^{2+}$-$H_2O$ states, which opens the Auger decay of the lowest C 2s vacancy in the pyrrole (Figure 4). However, our results show that the lowest ICD and Auger final states in imidazole-*d*-$H_2O$ are lying at the center of the C $2s^{-1}$ band (Figure 3), so there is some possibility that these channels are open. Doubts are related to the tendency of the ADC(2) method to lower the DIP energies, sometimes by up to about 0.5−1.0 eV,[44,45] and thus, both the ICD and Auger decay are more likely to be energetically closed. In contrast, the impact of the electron-density accepting water molecule in both imidazole-*a*-$H_2O$ (Figure 3a) and pyridine-*a*-$H_2O$ (Figure 4a) is such that the ICD channel is open for the C 2s ionized cation-radicals, while the Auger decay is closed.

Let us now turn to the processes that can be initiated by an inner-valence ionization of the water molecule, or by the population of the O $2s^{-1}$ states that in all complexes considered in this work lie between 31 and 34.5 eV. Here, an additional type of final states, which is not depicted in Figures 3 and 4, should be mentioned. Namely, the states with two vacancies on the water molecule, corresponding to the final states of an Auger decay within the water. It is known that the lowest doubly ionized state of a single water molecule has energy above 38 eV, which is shifted down to about 36.5 eV in water dimer for the electron-density accepting molecule.[30] In the molecular pairs of heterocycles with water considered in this work, there are also energy shifts of the dicationic states of water due to the donor or acceptor behavior of the neighboring heterocycle. In the cases in which the water molecule acts as electron-density acceptor (imidazole-*a*-$H_2O$ and pyridine-*a*-$H_2O$), the lowest state corresponding to two vacancies on the water appears at 34.3 and 34.4 eV, respectively, which is more than 1 eV higher than the O $2s^{-1}$ band (Figures 3a and 4a, lower panels). When water is in a donating position, the lowest $H_2O^{2+}$ state appears even higher in energy, reaching 39.7 and 39.5 eV in imidazole-*d*-$H_2O$ and pyrrole-*d*-$H_2O$, respectively. The possibility for a local Auger decay after inner-valence ionization of water is therefore energetically closed in all studied molecular complexes.

Instead, the O $2s^{-1}$ states can decay by ionizing the neighbor in an ICD or ETMD process to the large number of lower-lying doubly ionized states (Figures 3 and 4). As a result, the ETMD-electron spectra, coming solely from the decay of the water O $2s^{-1}$ vacancy, are quite broad. Moreover, when water acts as an electron-density donor, the O $2s^{-1}$ band shifts to higher energies, while the lowest states with two vacancies on the heterocycle shift to lower energies, which together insures a larger energy gap between the possible initial and final states of ETMD. This results in a more pronounced and broad ETMD-electron spectrum for imidazole-*d*-$H_2O$ compared to imidazole-*a*-$H_2O$ (Figure 3), as well as for pyrrole-*d*-$H_2O$ compared to pyridine-*a*-$H_2O$ (Figure 4).

According to the present results, the ICD and ETMD electrons can be expected to contribute substantially to the secondary-electron spectra of heterocycle−water complexes in a relatively wide energy interval. One can expect that this contribution will be even higher with more water molecules surrounding the heterocycle because of the additional nonlocal relaxation channels that will be opened between the water molecules[19] and involving the solvated molecule.[24]

Our results reveal that the location of the water molecule is critical for the nature of the lowest dicationic states. In the case of inner-valence ionization of the heterocycle, the system can undergo both Auger and ICD processes, while in the case of inner-valence ionization of the water only nonlocal relaxations, namely, ICD and ETMD, are possible. That is, one can expect that the inner-valence ionization of water in the vicinity of a biomolecule is likely to cause damage of the latter by inducing its single or even double ionization through an ICD or an ETMD mechanism. Considering the ionization of the heterocycle, the inner-valence vacancies located on carbon atoms are of particular interest since their relaxation strongly depends on the electron-density donating or accepting behavior of the water molecule. The presence of water in an electron-density accepting position destabilizes the heterocycle$^{2+}$−water states, final for the Auger process, while the stabilization of water$^{+}$ results in opening of the ICD channel for the lowest C $2s^{-1}$ state of the heterocycle. When water is located in an electron-density donating position, the screening of the positive charges leads to stabilization of the heterocycle$^{2+}$−water final states. The downshift in energy of the latter is so pronounced that in the presence of water, the Auger-decay channel opens for the lowest C $2s^{-1}$ state of pyrrole. It can be expected, therefore, that the relaxation of inner-valence vacancies located on the carbon atoms of large biomolecules is also dependent on the electron-density donating or accepting behavior of the neighbor, either water or another biomolecule.

## ■ COMPUTATIONAL METHODS

Optimal geometries of all the structures considered in this work, as well as the normal modes, and basis-set superposition errors (BSSE) corrections were obtained using the MP2/aug-cc-pVDZ level of theory, as implemented in the Gaussian16 program suite.[39]

The vertical single-ionization spectra of all the complexes considered were computed using the third-order algebraic diagrammatic construction[46−49] (ADC(3)) scheme (in its non-Dyson variant[50]) for approximating the one-particle Green's function. The ADC(3) method is applicable in situations where the breakdown of the molecular orbital picture of ionization[51] takes place. The latter phenomenon manifests itself by a strong redistribution of spectral intensity from the 1h main states to 2h−1p satellites and is typical for the inner-valence vacancy region considered in this work. The ADC(3) method was previously used in various applications, including computations of ionization spectra of valence-,[51,52] inner-valence-,[53−55] and K-shell[56] orbitals, and is proved to be a valuable tool for spectroscopic studies. The ionization spectra reported in this work were computed using the in-house ADC(3) code linked to the GAMESS-UK program suite[57,58] using cc-pVDZ basis sets. To account for the vibrational broadening and experimental resolution, the resulting spectra were convoluted with Gaussians with fwhm (full width at half-maximum) of 0.4 eV.

The double-ionization spectra of the complexes were computed at the ADC(2) level of theory for obtaining the particle−particle propagator[59,60] with cc-pVDZ basis sets, again using our in-house computer program. The character of the final doubly ionized states with respect to the location of the vacancies (holes) was determined by analyzing the leading wave function amplitudes for each state using the two-hole population analysis proposed in ref 61. The final ICD states are characterized by vacancies residing on different molecules, while in the final Auger and ETMD states both vacancies are located on the same molecule. The difference between the latter processes is the initial vacancy, which is localized on the same molecule, in the case of an Auger transition, or on the different one, in the case of an ETMD process.

Based on the calculations described above, the electron kinetic energy distribution spectra can also be modeled, as described in ref 30, by subtracting the energy of each individual line of the DIP-spectrum from each of the lines of the IP-spectrum which is energetically open for decay into this particular final dicationic state. The probability of each transition is evaluated with respect to the populations of the initial and the final states, assuming that it is independent of the energy differences between the states. The resulting spectra were convoluted with a Gaussian having a fwhm of 0.5 eV. Such a simplified approach has proven to be quite effective when compared to experimental data.[27]

## ■ AUTHOR INFORMATION

### Corresponding Author

**Anna D. Skitnevskaya** − *Theoretische Chemie, Physikalisch-Chemisches Institut, Universität Heidelberg, Heidelberg 69120, Germany; Laboratory of Quantum Chemical Modeling of Molecular Systems, Irkutsk State University,*

664003 Irkutsk, Russia; orcid.org/0000-0002-9328-1598; Email: a.skitnevskaya@isu.ru

### Authors

**Kirill Gokhberg** − *Theoretische Chemie, Physikalisch-Chemisches Institut, Universität Heidelberg, Heidelberg 69120, Germany*

**Alexander B. Trofimov** − *Laboratory of Quantum Chemical Modeling of Molecular Systems, Irkutsk State University, 664003 Irkutsk, Russia; Favorsky's Institute of Chemistry, SB RAS, 664033 Irkutsk, Russia*

**Emma K. Grigoricheva** − *Theoretische Chemie, Physikalisch-Chemisches Institut, Universität Heidelberg, Heidelberg 69120, Germany; Laboratory of Quantum Chemical Modeling of Molecular Systems, Irkutsk State University, 664003 Irkutsk, Russia*

**Alexander I. Kuleff** − *Theoretische Chemie, Physikalisch-Chemisches Institut, Universität Heidelberg, Heidelberg 69120, Germany*

**Lorenz S. Cederbaum** − *Theoretische Chemie, Physikalisch-Chemisches Institut, Universität Heidelberg, Heidelberg 69120, Germany;* orcid.org/0000-0002-4598-0650

Complete contact information is available at:
https://pubs.acs.org/10.1021/acs.jpclett.2c03654

### Notes

The authors declare no competing financial interest.

## ■ ACKNOWLEDGMENTS

A.D.S., E.K.G., A.I.K., and L.S.C acknowledge the financial support by the European Research Council (ERC) (Advanced Investigator Grant 692657). A.B.T. acknowledges the grant, Grant No. FZZE-2020-0025, from the Ministry of science and higher education of the Russian Federation.

## ■ REFERENCES